\documentclass[pre,floats,twocolumn,showpacs,superscriptaddress]{revtex4}

\usepackage{graphicx,epsfig}
\usepackage{dcolumn,bm,fleqn,epic,eepic,float}
\usepackage{amssymb,amsmath,multirow,rotate,color}
\usepackage{subfigure}
\usepackage{hyperref}

\begin{document}

\title{Velocity-enhanced Cooperation of Moving Agents playing Public Goods Games}

\author{Alessio Cardillo}
\affiliation{Instituto de Biocomputaci\'on y F\'{\i}sica de Sistemas Complejos, Universidad de Zaragoza, E-50018 Zaragoza, Spain.}
\affiliation{Departamento de F\'{\i}sica de la Materia Condensada, Universidad de Zaragoza, E-50009 Zaragoza, Spain.}

\author{Sandro Meloni}
\affiliation{Instituto de Biocomputaci\'on y F\'{\i}sica de Sistemas Complejos, Universidad de Zaragoza, E-50018 Zaragoza, Spain.}

\author{Jes\'us G\'omez-Garde\~nes}
\affiliation{Instituto de Biocomputaci\'on y F\'{\i}sica de Sistemas Complejos, Universidad de Zaragoza, E-50018 Zaragoza, Spain.}
\affiliation{Departamento de F\'{\i}sica de la Materia Condensada, Universidad de Zaragoza, E-50009 Zaragoza, Spain.}

\author{Yamir Moreno}
\affiliation{Instituto de Biocomputaci\'on y F\'{\i}sica de Sistemas Complejos, Universidad de Zaragoza, E-50018 Zaragoza, Spain.}
\affiliation{Departamento de F\'{\i}sica Te\'orica, Universidad de Zaragoza, E-50009 Zaragoza, Spain.}

\date{\today}

\begin{abstract}
In this paper we study the evolutionary dynamics of the Public Goods Game in a population of mobile agents embedded in a $2$-dimensional space. In this framework, the backbone of interactions between agents changes in time, allowing us to study the impact that mobility has on the emergence of cooperation in structured populations. Our results point out that a low degree of mobility enhances cooperation in the system. In addition, we study the impact of the size of the groups in which games are played on cooperation. Again we find a rise-and-fall of cooperation related to the percolation point of the instant interaction networks created by the set of mobile agents.
\end{abstract}
\maketitle

%

Despite its ubiquity in nature and human societies, the survival of cooperative behavior among unrelated agents (from bacteria to humans) when defection is the most advantageous strategy is not fully understood and constitutes one of the most fascinating theoretical challenges of Evolutionary Theory \cite{axelrod,maynard}. Recently, it has been pointed out that the integration of the microscopic patterns of interactions among the agents composing a large population into the evolutionary setting provides a way out for cooperation to survive in paradigmatic scenarios such as the Prisoner's Dilemma (PD) game \cite{nowak,szaborev,anxorev}. 
The structural features studied span from simple regular lattices \cite{nowak93,arenzon2001,sss08} to real patterns displayed by social networks \cite{rev:bocc}, such as the small-world effect \cite{abramson}, scale-free (SF) patterns for the number of contacts per individual \cite{PRLsantos,PNASsantos,PRL,JTB,tang06}, the presence of clustering \cite{push,assenza} or modularity \cite{arenas}. 

Although the above studies mostly focus on the PD game, other paradigmatic settings have also been studied on top of network substrates, such as the Public Goods Game. The Public Goods Game (PGG) is seen as the natural extension of a PD game when passing from pairwise to $n$-person games. 
After the work by Santos {\em et al.} \cite{nature} showing that SF architectures promote cooperation, many other works have continued this line of research by exploring the networked version of the PGG \cite{szolnoki,pgg1,pgg2,szolnoki2,pgg3,pgg4,perc}. Moreover, as the PGG formulation introduces two structural scales, namely individuals and the groups within which they interact, it has been shown that the structure of the mesoscale defined by the groups also play an important role for the success of cooperation \cite{chaos,epl,szolnoki3,perc2}.

The assumption of a static graph that maps social ties, although still a coarse grained picture of the microscopic interactions, provides with a useful approach for studying the dynamics of large social systems. However, when moving to smaller scales one has to consider additional microscopic ingredients that may influence the collective outcome of social dynamics. One of these ingredients is the mobility of individuals, a topic that has recently attracted a lot of attention, and that has been  tackled from different perspectives.  The range of studies in which mobile agents have been included spans from pure empirical studies \cite{batty_barthelemy_plosone_2011,barrat-plosone-2011, barrat-ad-hoc-networks-2011}, to theoretical ones that focus on the role that mobility patterns have on different dynamical processes such as disease spreading \cite{frasca_pre_2006}, synchronization \cite{frasca_prl_2008, albert-dg_pre_2011} and evolutionary dynamics \cite{meloni_pre_2009,zhang_physa2011,aktipis04,vainstein07,helbing_pnas_2011, cheng_njp_2010} in the context of the PD game. In addition, more complex representations in which an entanglement between agents mobility and evolutionary dynamics is introduced have been studied within the framework of the PD game  \cite{helbing08,helbing_pnas_2009,helbing_ejpb09,cheng_njp_2011} and, more recently, in the context of the PGG \cite{aktipis11}. 

In this Brief Report we follow the setting introduced in \cite{meloni_pre_2009} in which a population of $N$ agents moves on a $2$-dimensional space. Simultaneously to the  movement of the agents we consider that a PGG is played. To this end, the movement dynamics is frozen at equally spaced time steps and each node engages its closest neighbors to participate in a group in which a PGG is played. Obviously, the mobility of individuals turns the usual static backbone of interactions into a time-evolving one, opening the door to novel effects on the evolution of cooperation. 
Our results point out a non-trivial dependence on the velocity of the agents and the group size in which PGG are played, yielding optimum operation points at which cooperation is favored. 


We start by introducing the dynamical setting in which the evolutionary dynamics of the PGG is implemented. 
Our population is composed of a set of $N$ agents living in the area inside a square with side length $L$. Thus, the density of individuals is defined as $\rho=N/L^2$. Both the density and the number of agents remain constant along our simulations. Our agents are initially scattered at random on top of the surface by using two independent random variables uniformly distributed in $[0,L]$ for assigning the initial position $[x_i(0),y_i(0)]$ of each agent. 

Once the initial configuration of the system is set, two dynamical processes co-evolve: movement and evolutionary dynamics. At each time step $t$, the movement of agents affects their current positions, $[x_i(t),y_i(t)]$ with $i=1,...,N$, by means of the following equations:
\begin{eqnarray}
x_i(t+1) =& \!x_i(t) + v\cdot \cos\theta_i(t) \;,
\label{eqx}
\\ 
y_i(t+1) =& \!y_i(t) + v\cdot \sin\theta_i(t) \;.
\label{eqy}
\end{eqnarray}
The value of each angular variable, $\theta_i$, is randomly assigned for each agent at each time step from a uniform distribution in the interval $\left[-\pi , \pi\right]$. In addition to the above equations, we use periodic boundary conditions so if one agent reaches one side of the square, it re-appears on the opposite one. 

The second ingredient of the dynamical model is the evolutionary PGG played by the mobile agents. In addition to the random assignment of its initial position, each agent is assigned its initial strategy randomly, so that with equal probability an agent is set as Cooperator [$s_i(0)=1$] or Defector  [$s_i(0)=0$].
After this initial stage, both movement and evolutionary dynamics evolve simultaneously. At each time step, just after each agent has updated its position in the plane as dictated by Eqs. (\ref{eqx}) and (\ref{eqy}), agents play a round of the PGG as follows. First a network of contacts is constructed as a Random Geometric Graph (RGG) \cite{rev:bocc}. Each pair of agents, $(i,j)$, creates a link between them provided they are separated less than a certain threshold distance, R: $\sqrt{(x_i(t)-x_j(t))^2+(y_i(t)-y_j(t))^2}\leq R$. After all the nodes have stablished their connections with their nearest neighbors, a RGG for the network of contacts at time $t$ emerges, whose topology is encoded in an adjacency matrix, $A_{ij}^t$, with entries $A_{ij}^t=1$ when nodes $i$ and $j$ are connected at time $t$ and $A_{ij}^t=0$ otherwise.

Once the RGG is formed, each of the agents defines, together with her $k_i$ nearest neighbors in the RGG, a group of size $k_i+1$  in which one PGG is played. 
In each of the groups she participates in, a cooperator player contributes an amount $c$ while a defector does not contribute. Besides, the total contribution of a group is multiplied by an enhancing factor $r$ and distributed equally among all the participants. Thus the total payoff accumulated by an agent $i$ at time $t$ reads:
\begin{equation}
P_i(t)\!=\!\!\sum_{j=1}^{N}(A^t_{ij}+\delta_{ij})\frac{\sum_{l=1}^{N}(A^t_{jl}+\delta_{jl})s_{l}(t)cr}{k_{j}(t)+1}-[k_i(t)+1]s_{i}(t)c\;.
\end{equation}

\begin{figure}[t!]
\centering
\includegraphics[width=0.8\columnwidth]{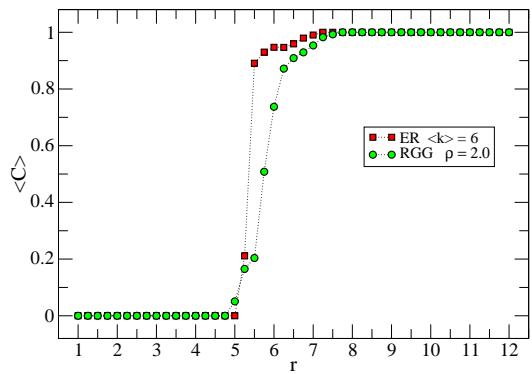}
\caption{(color online) Average fraction of cooperators $\langle c \rangle$ with respect to the enhancement factor $r$ for\emph{static} RGG and ER networks. Both networks have the same number of nodes $N=1000$ and average degree $\langle k \rangle = 6$.}
\label{fig:static-networks}
\end{figure}

After each round each of the agents can update her strategy. To this aim, an agent $i$ chooses one of her instant neighbors $j$ at random and with probability $\Pi[s_i(t+1)=s_j(t)]$, $i$ will take the strategy of $j$ during the next round of the PGG. The former probability reads:
\begin{equation}
\Pi[s_i(t+1)=s_j(t)]=\frac{\Theta[P_j(t)-P_i(t)]}{M[k_i(t),k_j(t)]}\;,
\end{equation} 
where $\Theta(x)=x$ when $x>0$ while $\Theta(x)=0$ otherwise, and $M(k_i,k_j)$ is the maximum possible payoff difference between two players with instant degres $k_i(t)$ and $k_j(t)$. In our simulations, we let co-evolve both movement and evolutionary dynamics  during $5\cdot10^4$ time steps. We take the first $25\cdot10^3$ steps as a transient period while the degree of cooperation of the system is measured during the second half of the simulations as $\langle c\rangle=\sum_{t=\tau}^{\tau+T}\sum_{i=1}^Ns_i(t)/T$, with both $\tau=T=25\cdot10^3$. The results reported below are averaged over different realizations (typically $50$).

We start our analysis by considering the static case in which the velocity of the agents is set to $v=0$. In this case, the RGG is fixed from the initial configuration while only the strategies of agents evolve. A RGG is described by a Poissonian distribution, $P(k)=\langle k\rangle^k {\mbox e}^{-\langle k\rangle}/k!$, for the probability of finding a node connected to $k$ neighbors. This distribution corresponds to a homogeneous architecture in which the dispersion around the mean degree, $\langle k\rangle$, is rather small. The same pattern for the degree distribution $P(k)$ is obtained for the typical Erd\H{o}s-R\'enyi random network model. However, the main differences between RGG and ER networks relies on the clustering coefficient, {\em i.e.} the probability that two nodes with a common neighbor share a connection, and their diameter. While in the case of ER graphs clustering vanishes as $N\rightarrow\infty$, the geometric nature of RGG boosts the density of triads leading to a finite and large clustering coefficient at the expense of being longer than ER graphs (since redundant links used to increase clustering do not contribute to create short-cuts). These differences are found to be relevant for the synchronization of RGG compared to ER graphs \cite{ijbc}. 

The results of the above analysis are shown in Fig.~\ref{fig:static-networks} where we represent the dependency of the average level of cooperation in the system $\langle c \rangle$ with respect to the enhancement factor, $r$, for both RGG and ER graphs having the same number of elements $N$ and the same average degrees $\langle k \rangle$. As expected, for low values of $r$ defection dominates the system while for large $r$ cooperation prevails. Between these two asymptotic regimes the transition from defection to cooperation occurs ($5\leq r\leq 8$) pointing out slight differences between RGG and ER graphs. In this region we observe that ER networks promote cooperation slightly more than RGG for which the transition curve towards full cooperation goes slower. This result seems to contradict previous observations in the context of PD game \cite{assenza} in SF networks. However, this latter result is  related to the increase of heterogeneity when clustering is enlarged in SF networks. In our case, this effect is not present and, alternatively, clustering induces important differences between ER and RGG regarding the average path length. This quantity is shown to be much larger in RGG than in ER graphs, thus making more difficult the percolation of cooperation in the whole system. On the other hand, the onset of both transitions are roughly the same.

%
\begin{figure}[t!]
\centering
\includegraphics[width=0.67\columnwidth] {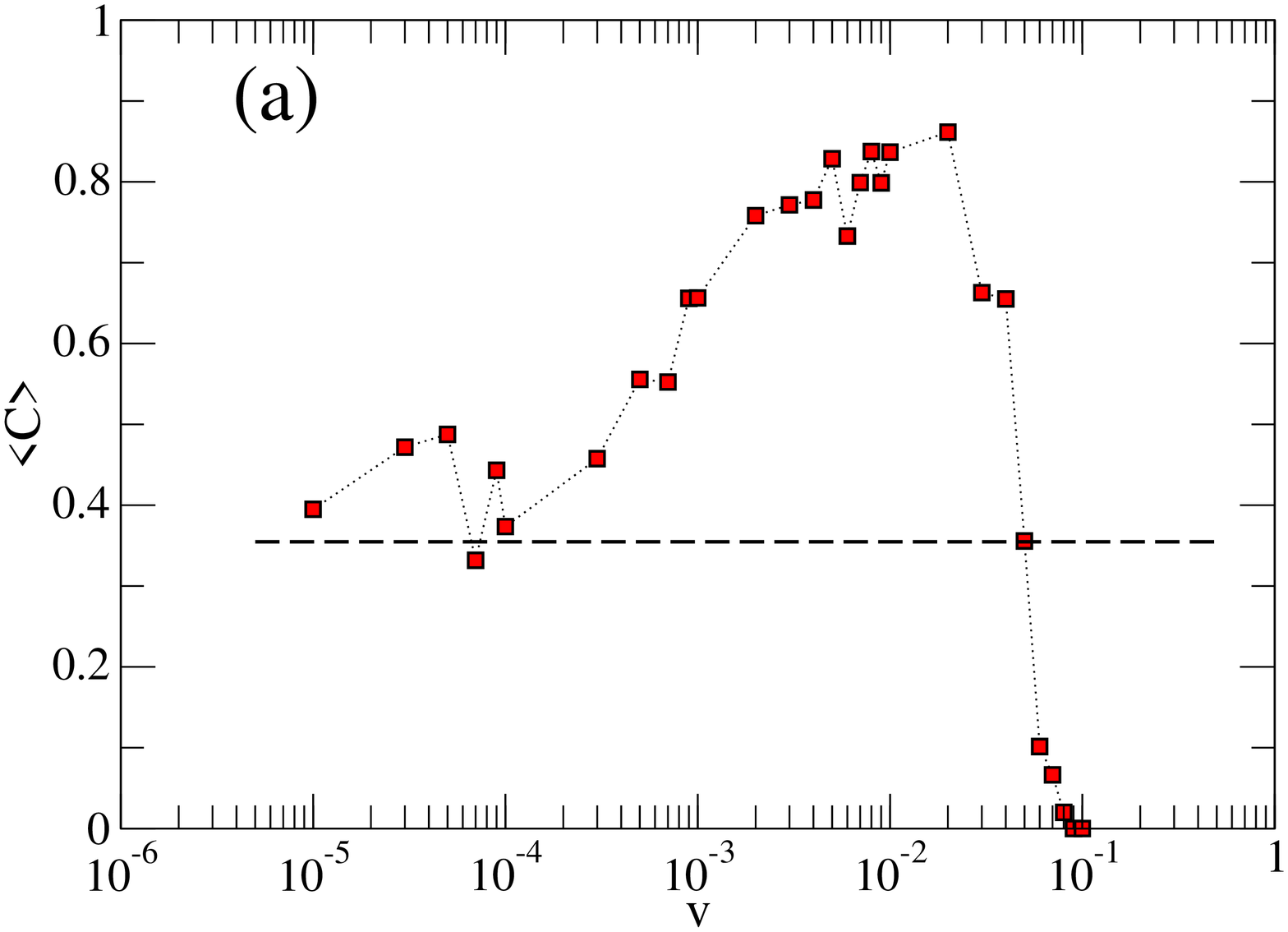}
\includegraphics [width=0.75\columnwidth] {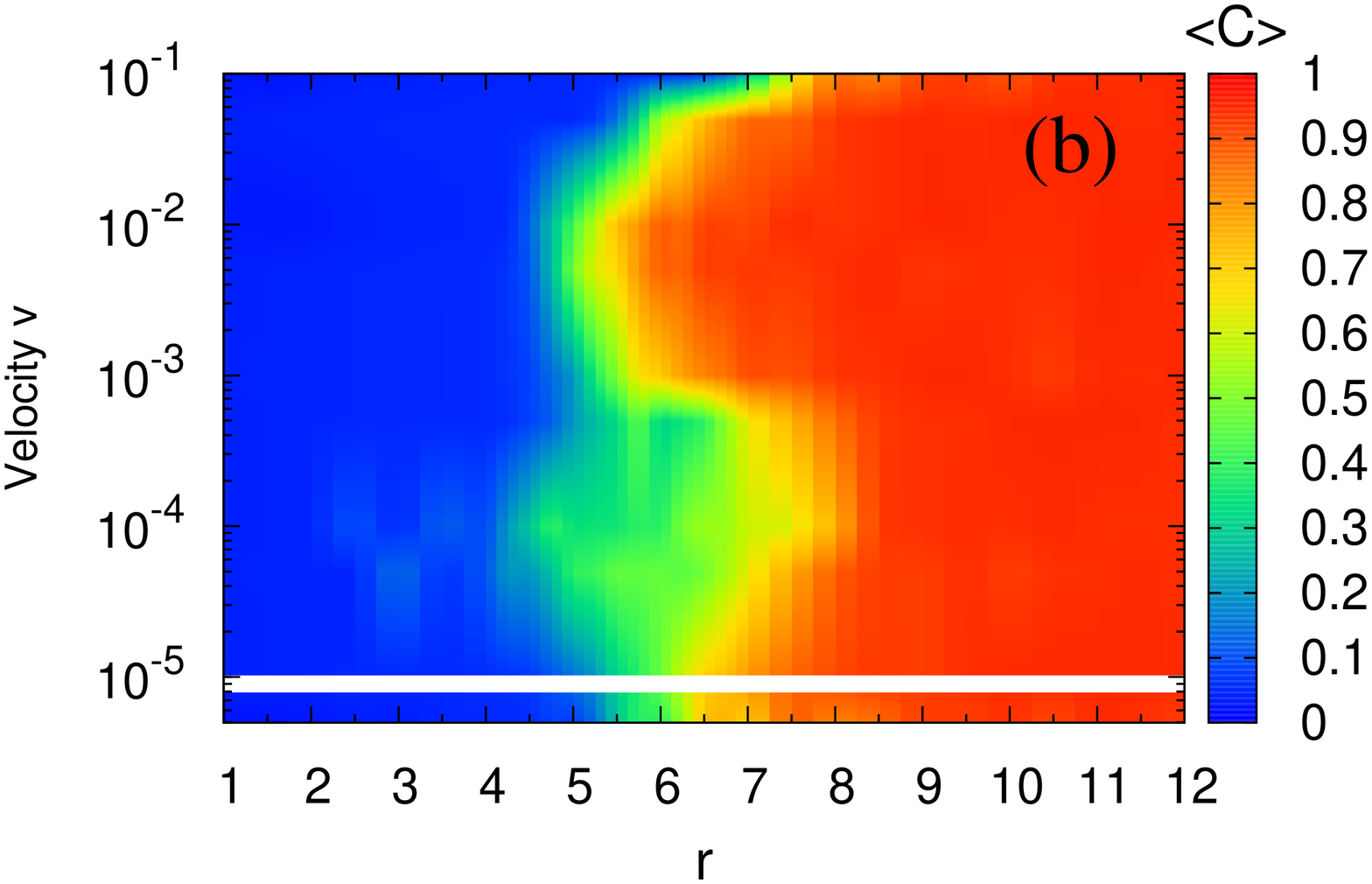}
\caption{(color online) Effects of velocity on the promotion of cooperation. (a) Average level of cooperation $\langle c \rangle$ as a function of the velocity $v$ of agents. The system has $\rho = 2.0$ and $R = 1.0$ and the enhancement factor $r$ is set to $r = 5.75$. The velocity spans in the interval $\left[10^{-5} ; 10^{-1} \right]$. The dashed line represents $\langle c \rangle$ in a static RGG with the same $N$ and $\langle k\rangle$, and for the same value of $r$. (b) Cooperation level $\langle c\rangle$ as a function of $v$ and $r$. The static case is displayed in the bottom strip of the panel (below the continuous thick band).}
\label{fig:picco-velocity}
\end{figure}
%
%

We now focus on the impact that the motion of agents has on the level of cooperation with respect to the static case. Thus, from now on, we consider that agents move with constant velocity $v$ following the rules given by Eqs. (\ref{eqx}) and (\ref{eqy}). Moreover, we set the value of the enhancement factor $r$ to be in the region for which the transition from full defection to cooperation occurs in the static case, namely $r=5.75$. Then, we monitor the degree of cooperation $\langle c\rangle$ in the system as a function of the velocity 
The results are shown in  Fig.~\ref{fig:picco-velocity}a together with the value (dashed line) for $\langle c\rangle$ in the static limit with $r=5.75$. We observe a rise-and-fall of cooperation so that when the velocity $v$ increases from very small values, the average level of cooperation increases significantly, reaching its maximum value for $v \simeq 2\cdot 10^{-2}$. From this point on, the increase of $v$ leads to the decay of cooperation so that $\langle c\rangle=0$ beyond $v \simeq 10^{-1}$. The fall of cooperation for large values of the velocity of agents is a quite expected result: as the velocity increases one approaches the well-mixed scenario for which cooperation is suppressed provided $r$ is less than the typical size of groups in which the PGG is played (here $\langle k\rangle=6$ so that groups are typically composed by $7$ agents). The rise of cooperation for small values of $v$ points out that there exists an optimal range for the velocity that allows a trade-off between two important ingredients for cooperator clusters to form and resist the invasion of defectors, namely: the ability to explore the plane to find other cooperators and a large enough time to interact with them so as to allow for the growth of cohesive cooperator clusters.

A more extensive analysis on the effects of motion is found in Fig.~\ref{fig:picco-velocity}b where a detailed exploration of the $(v,r)$-parameter space is shown together with the cooperation level in the static case (bottom part of the panel) as obtained from the corresponding curve in Fig.~\ref{fig:static-networks}. This panel confirms the results obtained in Fig.~\ref{fig:picco-velocity}a and provides a more complete picture about the enhancement of cooperation produced by the mobility of agents.  First, by comparing the bottom ($v=0$) and top ($v=10^{-1}$) parts of the panel, we observe that a large value of the velocity decreases the cooperation level of the static system. In particular, let us note that the transition region in the limit of large velocity is placed around $r\simeq 7$ thus recovering the well-mixed prediction. However, the relevant results are found between the static and large velocity limits. The effects of mobility in this region affect both the onset of the transition towards cooperation
and its fixation.
First, we observe that even for very low values of $v$ the onset of cooperation is anticipated with respect to the static case at the expense of having a broader transition towards full-cooperation as compared to the static RGG. However, when the velocity level is further increased, the transition becomes sharper and both the onset and the fixation of the full-cooperative state occur before with respect to the static case. 


Finally, we focus on the influence of group size in the  evolutionary success of cooperation. Considering a fixed velocity lying in the region for which the increase of cooperation is observed, namely $v=10^{-2}$, we compute the level of cooperation $\langle c\rangle$ as a function of the re-scaled enhancement factor, $\eta=r/(\langle k\rangle+1)$, where the denominator is the average size of the groups. This re-scaling is needed for the sake of comparing the cooperation levels for systems in which the group size is different. In this way, the well-mixed prediction is a sharp transition from full defection to full cooperation at $\eta=1$. As anticipated above, the size of the groups can be written as $\rho\pi R^2+1$, thus we can vary both the radius of interaction $R$ or the density of agents $\rho$. 

\begin{figure}[t!]
\centering
\includegraphics [width=0.8\columnwidth, angle=0] {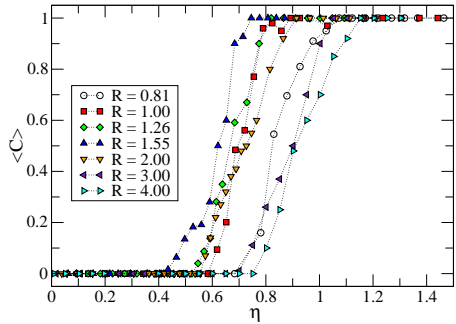}
\caption{(color online) Cooperation level $\langle c\rangle$ as a function of $\eta=r/(\langle k\rangle+1)$ varying the interaction radius of the agents ({\em i.e.} the size of the groups). In all the cases the density of players is $\rho = 2.0$ and their velocity is $v=10^{-2}$.}
\label{fig4}
\end{figure}

In what follows we vary the radius and keep the density constant to $\rho=2$. In Fig. \ref{fig4} we observe that, again, a rise-and-fall of cooperation is observed when going from low radius to large ones. Obviously, as radius (and hence group size) increases, we approach the well-mixed case so that the transition point reaches the theoretical value $\eta=1$. However, this approach is not monotonous and for intermediate  values of the radius and group size the cooperation transition is anticipated with respect to lower values of $R$. The reason behind this behavior relies on the percolation of the effective network when increasing the radius of influence of each agent. For low values of $R$ the effective network of contacts contains a number of disconnected clusters, however reaching the percolation threshold (meaning that the $\langle k\rangle=\rho\pi R^2>2$) nearly all the agents are incorporated into a macroscopic giant component. In our case $\rho=2$ so that the percolation point lies around $R\simeq\sqrt{\pi}\simeq1.773$ which agrees with the numerical observation in Fig. \ref{fig4}. This connection with the percolation point and the cooperation level has been recently observed in \cite{percsr} in the context of the PD game in regular lattices.

Summing up, the results presented in this Brief Report show that the mobility of the agents playing a PGG enhances cooperation provided their velocity is moderate. This enhancement is obtained by comparing the outcome of the evolutionary dynamics of the PGG with the results obtained in the static case. The addition of the random movement of agents produces the evolution in time of the original RGG, being the rate of creation and deletion of links controlled by the velocity of agents. When this rate is non-zero, allowing cooperators to explore the space, while moderate, so that cooperators clusters can be efficiently formed, we observe an optimal operation regime in which both the onset of cooperation and the fixation of cooperation in the system are enhanced. Finally, we have checked that group size shows a similar resonance phenomenon regarding the level of cooperation. However, in this case the point at which cooperation is enhanced is related to the percolation point of the effective network, {\em i.e.} with the point at which the size of the groups is large enough so as to have a macroscopic giant component for the network of contacts.

\acknowledgments
We acknowledge support from the Spanish DGICYT under projects FIS2008-01240, MTM2009-13848, FIS2009-13364-C02-01 and FIS2011-25167, and by the Comunidad de Arag\'on (FENOL). J.G.G. is supported through the Ram\'on y Cajal program.


\nocite{*}

\bibliographystyle{apsrev4-1}


\end{document}